\documentclass[aps,pre,reprint,showpacs,superscriptaddress]{revtex4-1}
\usepackage{amsmath}
\usepackage{amssymb}
\usepackage{gensymb}
\usepackage{graphicx}
\usepackage{epstopdf}
\usepackage[english]{babel}
\usepackage{siunitx}

\begin{document}

\title[Pump-probe XCCA studies of PtPOP dymanics]{Ultrafast Structural Dynamics of Photo-Reactions Revealed by Model-Independent X-ray Cross-Correlation Analysis}

\author{Peter Vester}
\affiliation{
Department of Physics, Technical University of Denmark, DK-2800 Lyngby, Denmark
}%

\author{Ivan A. Zaluzhnyy}
\altaffiliation[Present address: ]{Department of Physics, University of California San Diego, La Jolla, CA 92093, USA}
\affiliation{
	Deutsches Elektronen-Synchrotron DESY, Notkestra{\ss}e 85, D-22607 Hamburg, Germany
}%
\affiliation{
National Research Nuclear University MEPhI (Moscow Engineering Physics Institute), Kashirskoe shosse 31, 115409 Moscow, Russia
}

\author{Ruslan P. Kurta}
\affiliation{
	European XFEL, Holzkoppel 4, D-22869 Schenefeld, Germany
}%

\author{Klaus B. M\o{}ller}
\affiliation{
Department of Chemistry, Technical University of Denmark, DK-2800 Lyngby, Denmark
}%

\author{Elisa Biasin}
\affiliation{
	Department of Physics, Technical University of Denmark, DK-2800 Lyngby, Denmark
}%
\affiliation{
	PULSE Institute, SLAC National Accelerator Laboratory, CA 94025, Menlo Park, USA
}%

\author{Kristoffer Haldrup}
\affiliation{
	Department of Physics, Technical University of Denmark, DK-2800 Lyngby, Denmark
}%

\author{Martin Meedom Nielsen}
\email{mmee@fysik.dtu.dk}
\affiliation{
	Department of Physics, Technical University of Denmark, DK-2800 Lyngby, Denmark
}%

\author{Ivan A. Vartanyants}
\email{ivan.vartaniants@desy.de}
\affiliation{
	Deutsches Elektronen-Synchrotron DESY, Notkestra{\ss}e 85, D-22607 Hamburg, Germany
}%
\affiliation{
	National Research Nuclear University MEPhI (Moscow Engineering Physics Institute), Kashirskoe shosse 31, 115409 Moscow, Russia
}

\date{\today}

\begin{abstract}
We applied angular X-ray Cross-Correlation analysis (XCCA) to scattering images from a femtosecond resolution LCLS X-ray free-electron laser (XFEL) pump-probe experiment with solvated PtPOP ([Pt$_2$(P$_2$O$_5$H$_2$)$_4$]$^{4-}$) metal complex molecules. The molecules were pumped with linear polarized laser pulses creating an excited state population with a preferred orientational (alignment) direction. Two time scales of $1.9\pm1.5$~ps and $46\pm10$~ps were revealed by model-independent XCCA, associated with an internal structural changes and rotational dephasing, respectively. Our studies illustrate the potential of XCCA to reveal hidden structural information in a model independent analysis of time evolution of solvated metal complex molecules.
\end{abstract}

\maketitle

\section{Introduction}

Recent development of coherent X-ray sources, such as synchrotrons and X-ray free-electron lasers (XFELs), led to a substantial progress in time-resolved X-ray scattering techniques, which allows one to study structural dynamics on femtosecond scale, making it possible to track chemical reactions in real time \cite{Minitti2015}.
Despite a significant progress of X-ray scattering methods over the last decades, investigations of molecular structure and dynamics remains a challenging experimental task.
A general problem within the structural analysis framework of small- and wide-angle X-ray scattering (SAXS/WAXS) experiments from molecules in solution is to deduce a large number of structural parameters, including three-dimensional (3D) structural model of the molecules and their interactions with the surrounding solvent molecules, from a single azimuthally integrated one-dimensional (1D) scattering curve.
Moreover, the key structural parameters deduced from conventional SAXS/WAXS experiments are known to be strongly correlated with experimental parameters \cite{Haldrup2011}, which further complicates evaluation of molecular structure from the experimental data.

A possible way to enhance the structural information obtained in X-ray experiments is to excite molecules by a polarized pump laser which to a certain degree orients these molecules \cite{Lorenz2010} and utilize anisotropic information recorded by the two-dimensional (2D) X-ray detectors for better optimization of the structural models.
In this respect an angular X-ray cross-correlation analysis (XCCA) \cite{Wochner2009, Kurta2013a, Mendez2014, Kurta2015, Lehmkuhler2016, Zaluzhnyy2017a, Zaluzhnyy2017b} has a significant potential to extract and utilize anisotropic information contained in 2D diffraction patterns to provide additional constraints for structural models in the framework of conventional SAXS/WAXS analysis, and, importantly, reveal otherwise hidden information on structure and dynamics of molecules under investigation.

The method of angular intensity correlations in X-ray diffraction goes back to a pioneering work of Z. Kam \cite{Kam1977} and recently further developed in a number of publications (see for a review \cite{Kurta2016}).
This method can, in principle, reveal information about the structure of an individual particle in solution not available from azimuthally integrated SAXS/WAXS measurements.
Moreover, the angular distribution of scattered X-ray intensity also contains information about the spatial orientation of molecules.
The ultra-bright femtosecond X-ray pulses from XFELs provide an opportunity to measure scattering signals with a time resolution much higher than rotational relaxation times, enabling studies of molecular rotational dynamics by laser pump/X-ray probe experiments.
In this case, XCCA offers a model independent approach for investigation of structural dynamics of photo-excited ensembles of particles (molecules, proteins, etc) in solution.
In contrast to  conventional SAXS/WAXS techniques XCCA automatically separates scattering of bulk (isotropic) solvent from anisotropic solute in the experimental data.
Our studies show that XCCA may significantly enhance the information content of scattering images in systems of partially oriented solvated molecules.
The present work is a novel application of XCCA to investigate the structure and dynamics of solvated molecules.

This work is focused on the analysis of the experimental data obtained in the pump-probe X-ray scattering experiment performed at the Linear Coherent Light Source (LCLS)~\cite{Biasin2018,Haldrup2018}.
In contrast to these publications where the interpretation was based on a theoretical model of the molecular structure in solution, XCCA allowed us to study the molecular dynamics without any \textit{a priori} knowledge or assumptions (cosine squared distribution of photo-excited molecules, symmetric top shape of the molecules, etc.).
This work is a test bench to prove the validity of the XCCA technique to elucidate the structure and dynamics on molecular level.
To verify our findings we compare experimental results with simulated X-ray diffraction patterns obtained from density functional theory (DFT) calculations of molecular structure.

\section{Experiment}

\subsection{Photo-excitation of PtPOP molecules}

\begin{figure}
	\includegraphics[width = 0.8\columnwidth]{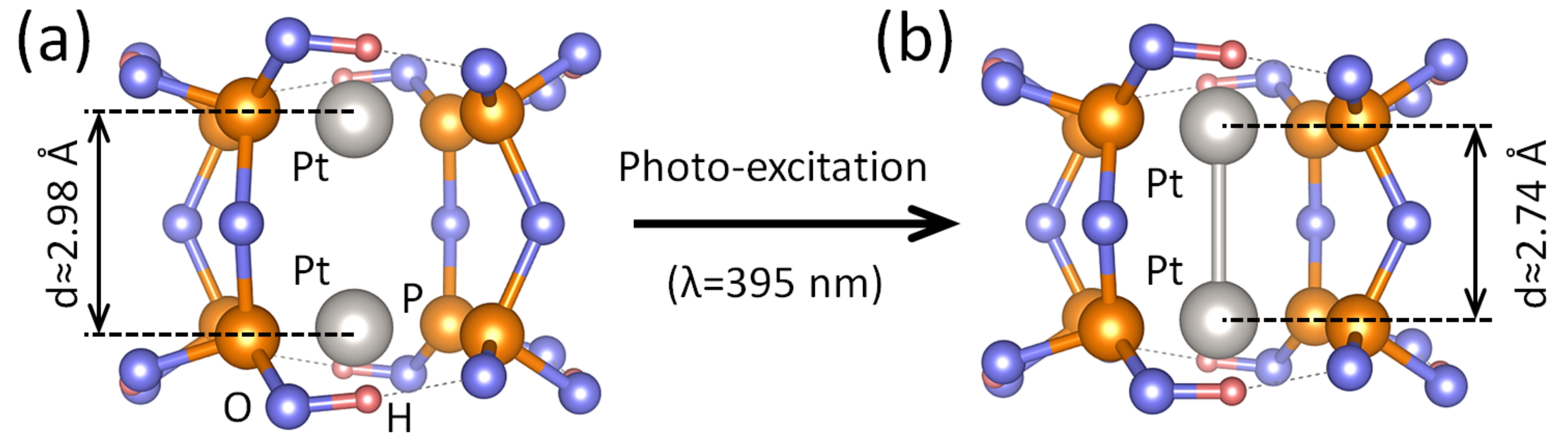}
	\caption{\label{PtPOP} (a) Structure of PtPOP molecule in ground state. (b) Structure of PtPOP molecule in excited state. Contraction of about 0.24~\AA~between two Pt atoms is shown.}
\end{figure}

The investigated metal complex molecules  tetrakis-$\mu$-pyrophosphitodiplatinate(II) anion ([Pt$_2$(P$_2$O$_5$H$_2$)$_4$]$^{4-}$, PtPOP) consist of a bi-planar Pt-Pt pair held together by four pyrophosphito ligands (Figure \ref{PtPOP}(a)).
It belongs to a family of binuclear $d^8$-$d^8$ transition metal complexes exhibiting photophysical properties of both fundamental and applied interest \cite{Gray2017} and shows intense luminescence with a total quantum efficiency very close to unity, a property which has made the system very amenable to investigation by time-resolved optical methods in both frequency and time domains.
It is now very well established, that upon photo-excitation at or near the 370 nm absorption peak, an electron is promoted from an anti-bonding $5d\sigma^\ast$ highest occupied molecular orbital (HOMO) to the bonding $6p\sigma$ lowest unoccupied molecular orbital (LUMO).
This chemical transition leads to a pronounced structural change in the form of contraction \cite{Christensen2009,VanderVeen2009} of the Pt atoms along the Pt-Pt axis by 0.24(4)~\AA~(Figure \ref{PtPOP}(b)).
Further, as the transition dipole moment lies along the Pt-Pt axis, the molecules will be selectively photo excited as a function of the Pt-Pt axis orientation relative to the polarization of the optical pump pulse \cite{Lorenz2010}, all of which makes it an excellent candidate for exploring the potential of XCCA (Figure \ref{Setup}(a)).

Following the photo-excitation event, the molecule is in a singlet state, and on a time scale of 1-10~ps undergoes inter-system crossing (ISC) to a triplet state.
The excited singlet- and triplet states, are well separated, both in their respective lifetimes (1-10~ps and 10~$\mu$s, respectively) as well as in terms of their potential energy surfaces, which are highly harmonic, nested potentials
shifted 0.24(4)~\AA~along the Pt-Pt coordinate with respect to the also highly harmonic ground state potential surface \cite{Gray2017, Levi2018}.
The quantum yield close to unity and well-separated optical signatures of the two states make PtPOP an almost ideal system for studying the fundamental phenomenon of non-radiative singlet-triplet transitions in the case where no intersections along the main reaction coordinate are immediately apparent.
However, to this day the intermediate state(s) and mechanism mediating the spin-state change remains elusive despite much recent work \cite{Lam2016, Monni2017, Monni2018}.
From these optical studies both direct interaction with the solvent, as well as deformation of the POP ligands, have been suggested as possible mechanisms.
However, the suggested intermediate states are optically "dark" and methods directly sensitive to structure are needed.

\subsection{Pump-probe experiment}

\begin{figure}
	\includegraphics[width=\columnwidth]{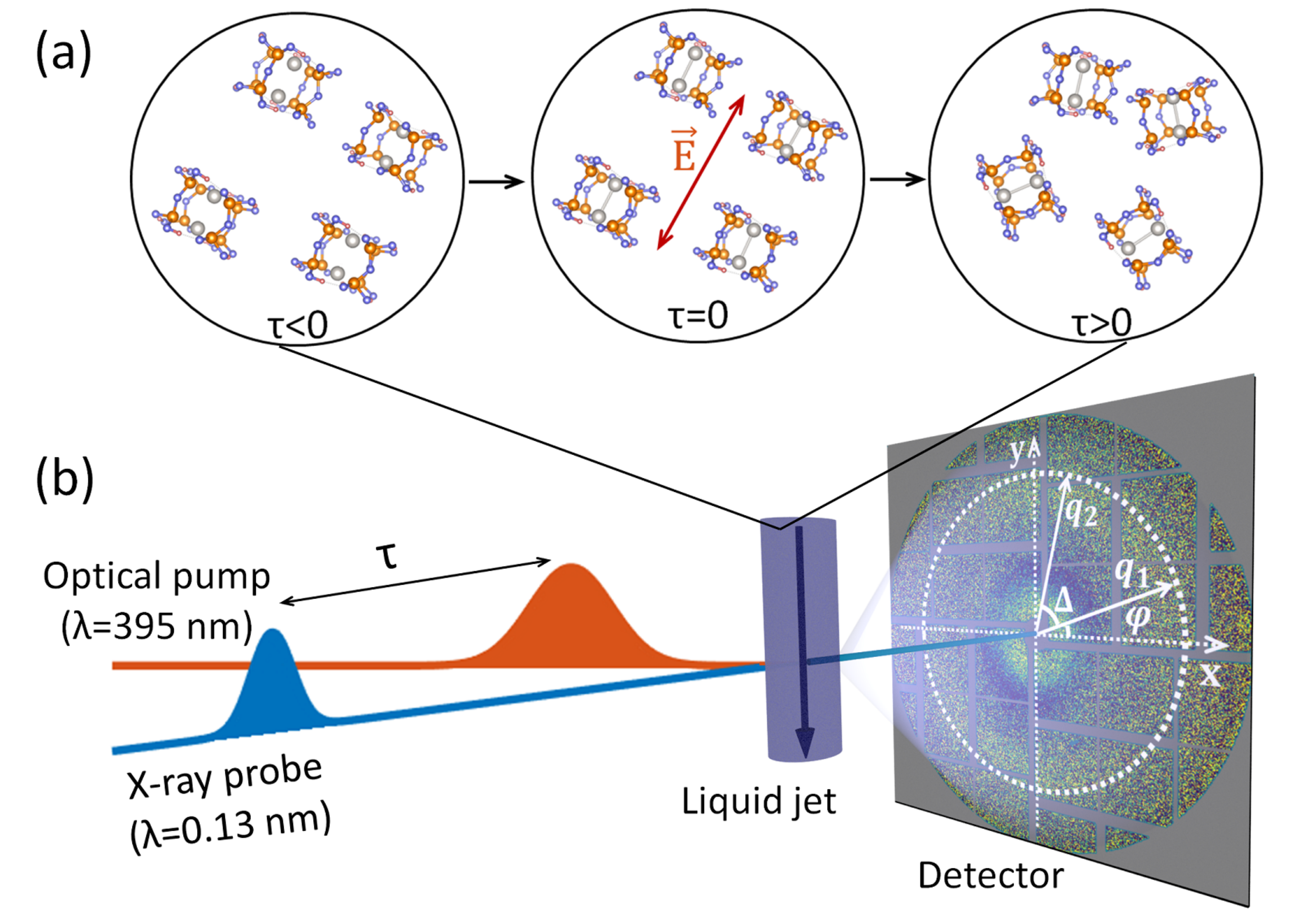}
	\caption{\label{Setup} (a) Temporal evolution of an ensemble of randomly oriented PtPOP molecules before ($\tau < 0$) and after ($\tau \geq 0$)
		excitation.
		The optical pump selectively excites PtPOP molecules with a dipole moment parallel to the laser electric field $\textbf{E}$ at $\tau=0$ and the population of excited molecules eventually evolves ($\tau>0$) to a random orientational distribution on a timescale of tens of picoseconds.
		(b) Scheme of the pump-probe experiment at LCLS.
		The experiment utilizes the optical pump laser/X-ray probe detection scheme on a circular liquid jet system with the time resolution given by the time delay $\tau$ of the femtosecond X-ray pulse.
		On the detector momentum transfer vectors $q_1$ and $q_2$ are shown with the angular coordinates $\varphi$ and $\varphi + \Delta$.}
\end{figure}

The dynamics following photo-excitation of aqueous PtPOP molecules was tracked in time-resolved pump-probe X-ray diffuse scattering (XDS) experiments at the XPP beamline of the LCLS XFEL facility as schematically illustrated in Figure \ref{Setup}(b) (for details of the experimental setup see \cite{Lemke2013,VanDriel2015}).
The investigated sample was a 80 mM aqueous solution of PtPOP in a vertical free flowing cylindrical liquid jet of 50 $\mu$m in diameter at a flow rate sufficient to fully replace the sample between successive pump-probe events (120 Hz).
The sample was excited by a short 50 fs 5 $\mu$J laser pulse with wavelength of 395 nm (pump) followed by a 50 fs 9.5 keV X-ray pulse (probe) at a well-defined time delay.
The nearly collinear laser beam with a circular spot size of approximately 50 $\mu$m and the X-ray beam with an estimated spot size of 30 $\mu$m at full width at half maximum (FWHM) were spatially and temporally overlapped at the sample position in the middle of the liquid jet.
The X-ray probe pulses were polarized in the horizontal direction, whereas the laser pump pulses had a linear polarization 20 degrees off the vertical.
Thus, a single pump-probe event gives a snapshot of the configuration after photo-excitation at a single time-delay $\tau$, and by combining snapshots at different time-delays the dynamics of the excited molecules can be followed.

XDS signals were recorded in the forward direction by the large-area 2D CS-PAD detector \cite{Hart2012} positioned 10 cm behind the sample and corrected for such effects as polarization, solid angle, absorption, background/dark image subtraction and outlier rejection as previously described \cite{VanDriel2015}.
The sensitivity of the diffraction patterns to structural changes in the sample is increased by considering difference scattering images, which are created by subtracting a laser-off (all molecules in the ground state) image from the nearest laser-on images (some molecules in the excited state) in the sequence of collected detector images.
The difference scattering images contain only a change in diffraction signal from ground- and excited-state PtPOP molecules and their interaction with solvent cage, while the constant background arising from solvent scattering cancels out.
To increase signal-to-noise ratio the collected diffraction patterns were temporally binned into 1~ps bins using the timing tool at LCLS, which allowed us to collect sufficiently many diffraction patterns ($M\approx3,000$) within each time bin for reliable XCCA analysis.
The measured difference scattering images averaged over 1~ps intervals of two different time delays $\tau=0-1$~ps and $\tau=9-10$~ps are shown in Figure \ref{Detector}(a,b).
One can see slight anisotropy in the intensity of the difference scattering images, parallel to the laser polarization (about 20 degrees to vertical direction), which appears due to the photo-excitation selectively occurring in molecules with Pt-Pt axis oriented along the polarization vector of the pump laser pulse \cite{Lorenz2010}.

\section{X-ray Cross-Correlation Analysis}

Angular anisotropy of difference diffraction patterns was analyzed by XCCA, which is based on evaluation of two-point angular cross-correlation functions (CCFs).
In this work, we apply the CCF defined on the scattering ring of radius $q$, where $\textbf{q}=(q,\varphi)$ is the momentum transfer vector defined in the polar coordinate system of the 2D detector \cite{Altarelli2010,*Altarelli2012, Kurta2016},
\begin{equation}
	\label{CCF}
	C(q,\Delta)=\langle I^{dif}(q,\varphi)I^{dif}(q,\varphi+\Delta)\rangle_\varphi \, ,
\end{equation}
where $I^{dif}(q,\varphi) = I^{On}(q,\varphi) - I^{Off}(q,\varphi)$ is the measured difference intensity between the laser on $I^{On}(q,\varphi)$ and laser off $I^{Off}(q,\varphi)$ diffraction patterns, $\Delta$  is the angular coordinate, and $\langle f(\varphi)\rangle_\varphi$ denotes the angular average of the function $f(\varphi)$.
In this work, the diffraction patterns were masked prior to the calculation of the CCF to exclude beamstop area, gaps between detector tiles  as well as not responding pixels on the detector. The pixels were also binned in $4\times4$ pixel groups to increase signal to noise ratio.

It is convenient to decompose the CCFs using an angular Fourier series on a ring of radius $q$
\begin{equation}
	\label{FCCF_1}
	C(q,\Delta)=\sum_{n=-\infty}^{\infty}C_n(q)e^{in\Delta} \, ,
\end{equation}
\begin{equation}
	\label{FCCF_2}
	C_n(q)=\frac{1}{2\pi}\int_{0}^{2\pi}C(q,\Delta)e^{-in\Delta}\mathrm{d}\Delta \, ,
\end{equation}
where $C_n(q)$ are the angular Fourier components of the CCF.
It can be shown~\cite{Altarelli2010,*Altarelli2012} that Fourier components of the CCFs are directly related to Fourier components of the difference intensities as
\begin{equation}
	\label{FCCF_5}
	C_n(q) =   \left| I_n^{dif}(q) \right|^2 \, .
\end{equation}

In practical applications, due to statistical variations of the CCFs, one needs to average the Fourier components \eqref{FCCF_5} over a large number $M$ (typically of the order of $10^3$ to $10^6$) of diffraction patterns to obtain reliable information about symmetry and structure of the system~\cite{Altarelli2010,*Altarelli2012}.
Averaged values of the Fourier components $\langle C_n(q) \rangle$ are directly related to the structure of molecules and their orientational distribution~\cite{Kurta2013b}.

\section{Results}
\subsection{Analysis of the angular anisotropy}

\begin{figure}
	\includegraphics[width = \columnwidth]{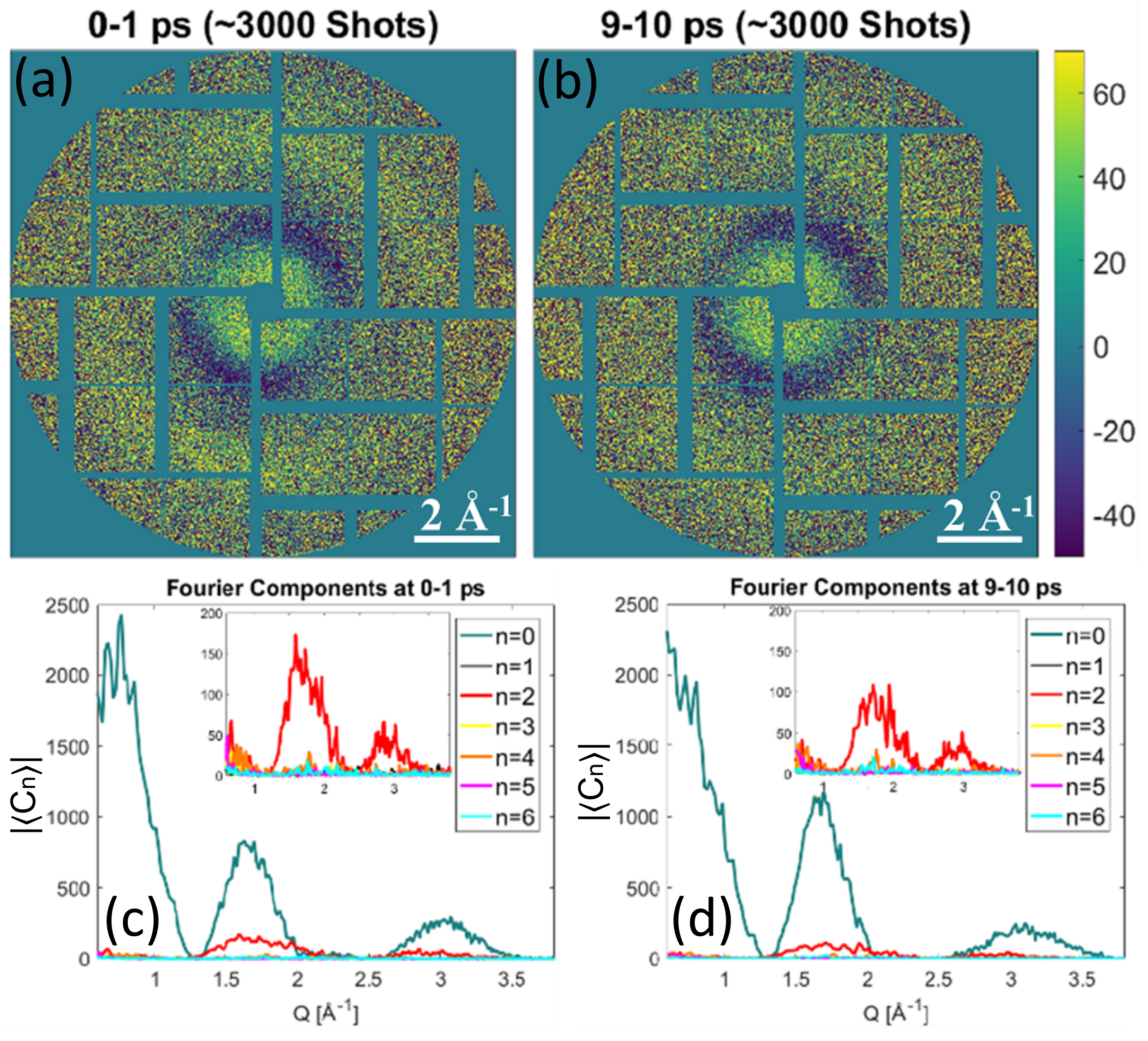}
	\caption{\label{Detector} (a-b) Difference scattering detector images (laser On - laser Off) for two different one picosecond time delay intervals.
		For better visualisation shown diffraction patterns were averaged over about 3,000 pulses within each time delay interval.
		(c-d) Calculated averaged Fourier components of the CCFs.
		The insets show the dominant anisotropic $n = 2$ Fourier component contribution (the isotropic $n = 0$ Fourier component is removed from the insets).}
\end{figure}

XCCA provides an excellent tool for model-independent analysis of angular anisotropy of difference diffraction patterns (Figure \ref{Detector}).
The averaged Fourier components of the angular CCF~\eqref{FCCF_5} from the experimental difference scattering intensities at two different time delays evaluated according to definition~\eqref{FCCF_2} are shown in Figure~\ref{Detector}(c,d).
According to equations \eqref{CCF}-\eqref{FCCF_2} the zero-order angular Fourier component ($n=0$) can be considered as the square of an azimuthally integrated difference intensity $I^{dif}(q,\varphi)$
\begin{equation}
	\label{FCCF_4}
	\langle C_0(q) \rangle \propto\Bigg\vert\int I^{dif}(q,\varphi)\mathrm{d}\varphi\Bigg\vert^2 \, .
\end{equation}
The first two peaks of $\left<C_0 (q)\right>$ at $q\approx0.7$~\AA$^{-1}$ and $q\approx1.8$~\AA$^{-1}$ (see Fig.~\ref{Detector}(c,d)) correspond to internal concentric rings with positive and negative signal on the difference scattering images in Fig.~\ref{Detector}(a,b).
A clearly visible peak of the same Fourier component in Fig. \ref{Detector}(c,d) at $q\approx3.0$~\AA$^{-1}$ (see Fig.~\ref{Detector}(c,d)) corresponds to the broad anisotropic external scattering ring in Figure~\ref{Detector}(a,b).

Information about angular anisotropy in diffraction can be conveniently accessed by evaluation of the higher-order Fourier components of the CCF $\left<C_n(q)\right>$, i.e. for $n=2,4,6,\ldots$.
A dominant $n=2$ angular Fourier component in the diffraction pattern is clearly visible in the inset of Figure \ref{Detector}(c,d).
A rapid decay of the dominant component after optical pump pulse is a strong evidence that the observed $n=2$ signal arises from a twofold symmetric orientational distribution of the excited state of the molecules induced by the laser excitation at time delay $\tau = 0$.
Weak but consistent $n=4$ and $n=6$ components can be also seen at around $q\approx1.8$~\AA$^{-1}$.
In principle, higher order intensity Fourier components in the X-ray scattering may originate from the internal symmetry of the individual molecules (see Figure \ref{PtPOP}(a)), which was directly observed in simulations with a relatively small number of illuminated particles \cite{Kurta2012}.

In general, the azimuthally averaged $n=0$ signal contains contributions from the solute, solvent cage and bulk solvent (heating and density changes caused by optical pump) \cite{Haldrup2010}.
The latter two should be subtracted from the azimuthally average data to extract the signal corresponding to the changes in solute structure \cite{Kjaer2013}.
In contrast, the $n=2$ signal is not influenced by the bulk solvent contributions but includes contributions from variations with the same symmetry as the orientational distribution of the molecules, e.g. solvent cage.
This significantly reduces the amount of free parameters in any structural analysis and makes the interpretation of the time constants more clear and requires less assumptions about the system.
A combination of both the $n=0$ and $n=2$ curves will give two different signals to benchmark the structural models, where one is dependent and one is independent of the bulk solvent contribution and therefore gives the possibility to enhance the structural information from a SAXS/WAXS experiment.

\subsection{Analysis of molecular dynamics}

\begin{figure}
	\includegraphics[width = \columnwidth]{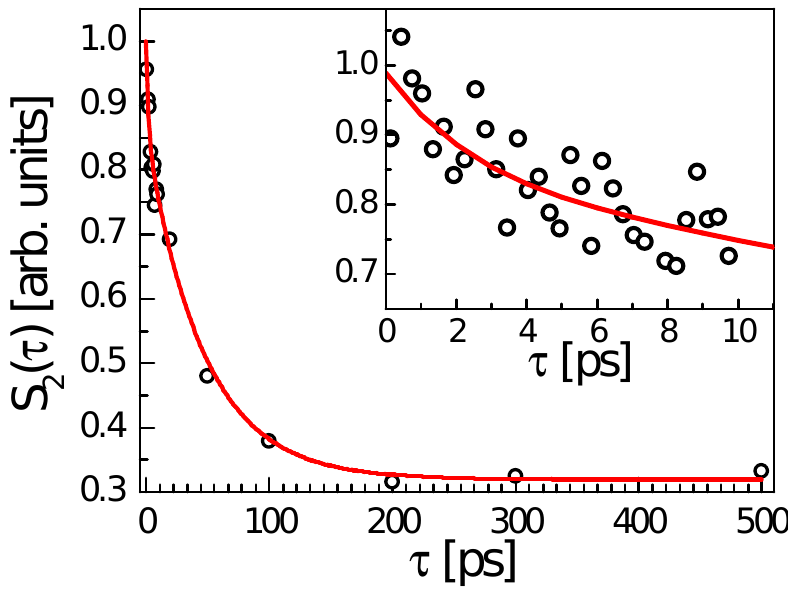}
	\caption{\label{Decay} A double exponential decay with the time constants $\tau_1=1.9\pm1.5$ ps and $\tau_2=46\pm10$ of the normalized area
		$S_2(\tau) = \int \sqrt{\langle C_2(q, \tau) \rangle} dq $ under the peak at $q=1.8$~\AA$^{-1}$.
		The average Fourier components of CCF $\langle C_2(q, \tau) \rangle$ were calculated according to equation~\eqref{FCCF_5} binned in 1~ps bins. In inset binning into 250~fs time bins reveals short time decay.
		Points experimental data, solid line fit to equation~\eqref{Exponential_decay}.}
\end{figure}

The values of non-zero CCF Fourier components contain all available information about the orientational distribution of photo-excited molecules.
To study the dynamics of excited PtPOP we considered the temporal evolution of the $n=2$ Fourier component.
To quantify the fraction of excited molecules within the cosine squared orientational distribution the normalized integrated area
$S_2(\tau) = \int \sqrt{\langle C_2(q, \tau) \rangle} dq$
under the peak at $q\approx1.8$~\AA$^{-1}$ of the averaged second-order Fourier component cross-correlation function
was considered (integration over $q$ was performed within the range 1.1 \AA$^{-1}$ - 2.3 \AA$^{-1}$).
The temporal evolution of this quantity as a function of delay time $\tau=0-500$~ps is shown by dots in Figure~\ref{Decay}.
One can clearly see a significant decay of the anisotropic signal after 10~ps.
Binning of the same data into 250 fs time bins, shown in the inset of Figure~\ref{Decay}, reveals an additional short-time decay, which is clearly visible on top of the longer decay.
Thus, the total decay was approximated by a sum of two exponential terms
\begin{equation}
	\label{Exponential_decay}
	S_2(\tau)=A e^{-\tau/\tau_1}+B e^{-\tau/\tau_2}+C \, ,
\end{equation}
where $A=21\pm5\%$ and $B=79\pm5\%$ are scale constants and $C$ is related to the general noise level of the averaged images.
Based on least-square fitting (see solid line in Figure~\ref{Decay}), the values of time constants were found to be $\tau_1=1.9\pm1.5$~ps and $\tau_2=46\pm10$~ps.
We would like to stress here that the values of time constants were obtained without any modeling or \textit{a priory} knowledge of system behavior.

The longer time scale $\tau_2=46\pm10$~ps may be interpreted as the rotational dephasing of the initial cosine squared distribution of orientations to a completely random and isotropic distribution.
The reorientation time $\tau_r$ for a molecule in solution can be estimated from the Stokes-Einstein-Debye hydrodynamic theory in a classical dynamical framework without electrical interactions as $\tau_r\approx50$~ps \cite{Biasin2018}.
The specific molecular shape of PtPOP can be taken into account (here we approximated  the shape of PtPOP molecule by a sphere with the radius $r=4$~\AA), which would lead to a slight change of the value of rotational time constant $\tau_r$.
If the laser excitation involves a change in size, shape or dielectric properties of a molecule, a different reorientation time can be expected for the excited species.
It means that the $n=2$ contribution from the ground and excited states can be distinguished by following the temporal evolution of the $n=2$ signal as the molecules with a longer time constant will dominate the signal at longer time delays.
This applies as well to experiments where different species (or same species of different sizes) in the same sample volume are excited with a preferred orientational direction.
The molecular reorientation time obtained from the $n=2$ signal can not be determined from radially integrated $n=0$ curves and can be used to refine structural models and gain new insight in time-resolved SAXS/WAXS experiments.

A direct interpretation of the shorter time constant $\tau_1=1.9\pm1.5$~ps is more challenging as it is assumed to arise from the internal dynamics of the molecule on short time scales.
The area under the two dominant peaks at $q\approx1.8$~\AA$^{-1}$ and $q\approx3.0$~\AA$^{-1}$ in the measured $n=2$ signal decreases with a similar time constant indicating that the shape of the signal remains almost unchanged.
Therefore, we interpret the short time constant as possibly reflecting the time scales for vibrational decoherence of the PtPOP molecule \cite{VanderVeen2011, Levi2018, Monni2018}, but the underlying mechanism is at present unknown.

\subsection{Model of the scattering signal}

\begin{figure}
	\includegraphics[width = \columnwidth]{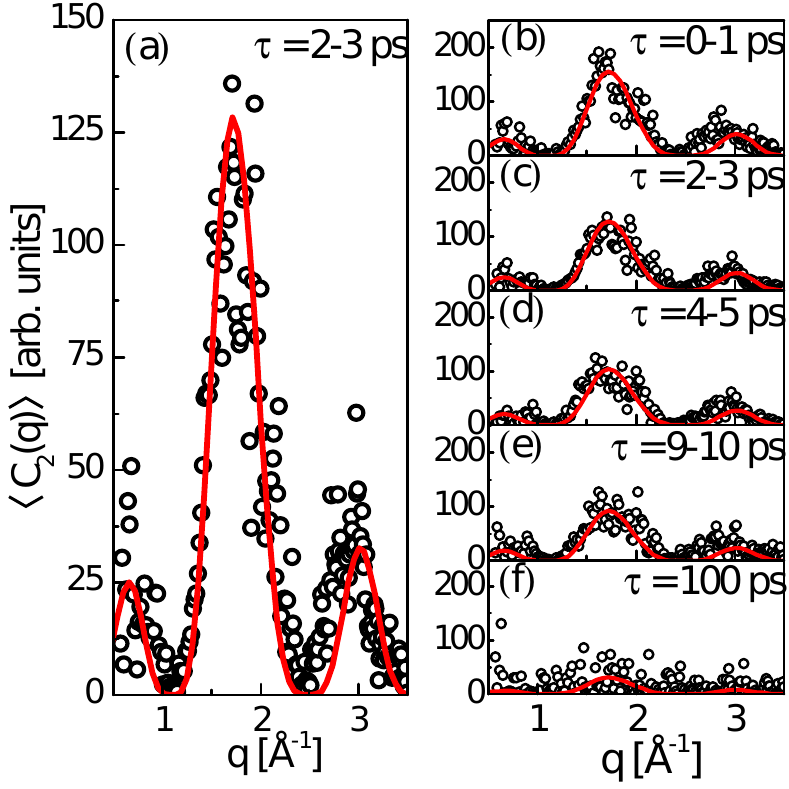}
	\caption{\label{Model} (a) Direct comparison of the calculated $n=2$ CCF Fourier components from the simulation of $10^5$ PtPOP molecules with a
		cosine squared angular distribution, fitted to the measured components in the $2-3$~ps time delay interval with a scaling factor $\alpha(\tau)$.
		(b-e) Same comparison for different time delays;
		at $\tau=100$~ps (f) the molecules have almost completely lost their preferred orientation.
		Here dots are experimental data obtained from XCCA analysis, solid lines are theoretical fit.}
\end{figure}

The time-dynamics results of XCCA can be directly compared to simulations based on a DFT structural model.
In this work we utilize a simple model for the excited state population of PtPOP, where they are considered as linear molecules.
Such model can be justified by rotational symmetry of PtPOP molecule around the Pt-Pt axis and the fact that the excitation of the molecule can be approximated to a high accuracy by Pt-Pt bond contraction.
When such symmetric top molecules are excited from thermal equilibrium by one-photon absorption, the orientational distribution of excited molecules will have a cosine squared distribution with respect to the polarization of the incoming optical photons \cite{VanKleef1999},
which was indeed observed in the collected X-ray diffraction patterns (Figure \ref{Detector}).

As a model system for simulation we assume a 3D disordered sample consisting of $N = 10^5$ molecules.
In the approximation of a dilute disordered sample where the mean distance between the molecules is larger than the coherence length of the incoming beam, interference between the X-rays scattered from different molecules can be neglected and the total scattered intensity can be represented as a sum of intensities from the individual molecules in the system.
The X-ray intensity scattered from one molecule can be evaluated as
\begin{equation}
	\label{Molecular_form_factor}
	I_{mol}(\textbf{q})=\Big\vert \sum_{i}f_i(q)e^{i \textbf{q} \textbf{r}_i} \Big\vert^2 \, ,
\end{equation}
where $\textbf{r}_i$ are the atomic positions and $f_i(q)$ are the atomic form factors of the i-th atom in the PtPOP molecule.

The coordinates of atoms in excited and ground state PtPOP structures were calculated by DFT simulations (see for details~\cite{Dohn2017, Levi2018}).
This gives a Pt-Pt bond contraction of 0.24~\AA,
while the ligand cage structure remains rigid, which is in agreement with experimental X-ray scattering results \cite{Christensen2009,VanderVeen2009}.

To obtain the diffraction patterns from an ensemble of molecules in excited state, we simulated our sample as one in which each molecule was rotated within the cosine squared distribution.
The positions of the atoms in rotated photo-excited PtPOP molecule were used to calculate a diffuse X-ray scattering signal from a single molecule using equation~\eqref{Molecular_form_factor}.
Then the diffraction signal was averaged over $N=10^5$ molecules and a difference scattering signal was calculated  to obtain similar diffraction patterns as we observed in the XDS experiment.

Figure~\ref{Model} shows a comparison of the experimentally observed and simulated $n=2$ Fourier component of the CCF at different time delays.
The model signal was scaled with a factor $\alpha(\tau)$ to account for the total number of excited PtPOP molecules in the probed volume of the sample and their orientational distribution at the specific time delay.
The scaling factor $\alpha(\tau)$ has a similar temporal decay behavior as observed for the $S_2(\tau)$ factor shown in Figure \ref{Decay}, with two time constants of $1.9\pm1.4$~ps and $41\pm8$~ps.

In this work we assumed that excited singlet and triplet states of PtPOP molecule are characterized by almost the same Pt-Pt distance within the accuracy of 0.01~\AA\
\cite{Stiegman1987, Levi2018}, which means that we could not observe singlet-to-triplet ISC in our pump-probe experiment.
Therefore, the temporal evolution of the scaling coefficient $\alpha(\tau)$ can be attributed only to the orientational dephasing of the excited molecules and not to the relaxation of the molecules to ground state, since the lifetime of triplet excited state is estimated to be about 10~$\mu$s.
A direct comparison confirms the validity of the assumed model of the system, and the small discrepancies are interpretated as arising from the lack of a simulated solvent cage signal and slight variations between the DFT simulated structure and the actual structure of the molecule.
This demonstrates how the direct structural modeling of these signals is possible in a straightforward and robust manner.

\section{Discussion}

In this work by applying XCCA we show that the difference signal from photo-excited PtPOP molecules can be well represented by contribution of two (zero- and second-order) Fourier components.
The dominant anisotropic signal of the $n=2$ Fourier component of the CCF clearly indicates that orientation of the photo-excited PtPOP molecules can be approximated by a cosine squared distribution.
This is in agreement with theoretical predictions \cite{Baskin2006, Lorenz2010}, assuming a single-photon excitation and initially non-occupied rotational and vibrational degrees of freedom of the PtPOP molecules \cite{Biasin2018}.
Observation of small values of higher-order Fourier components, indicates the possibility of XCCA to go beyond the common assumption about cosine squared distribution of molecular orientations and gain new insight on the structure of molecules.

Time-dependent analysis of the anisotropic scattering signal reveals that its shape remains unchanged, while the amplitude exponentially decreased to the noise level with two characteristic time-scales $\tau_1 = 1.9\pm1.5$~ps and $\tau_2 = 46\pm10$~ps.
Taking into account the long lifetime of photo-excited state of PtPOP molecules and the fact that the ISC from the singlet to the triplet state occurs with near-unity efficiency, this decay can be attributed exclusively to rotational dephasing of molecules (longer time constant) and internal dynamics of molecules (shorter time constant).
Our analysis is supported by simulation of the difference scattering signal, which shows that the anisotropic scattering can be modelled by difference scattering signal from excited- and ground-state molecules at any time-delay.
In principle, it should be possible to observe oscillations on a sub-300~fs timescale in the difference scattering signal, which is directly related to the Pt-Pt bond stretching mode \cite{VanderVeen2011,Levi2018}.
The direct studies of the bond dynamics would require collecting significantly more scattering patterns with short time delays to accumulate sufficient statistics for the XCCA.
This opens up the possibility to investigate, for example, the optical Kerr effect \cite{Palese1994,Jimenez1994,Castner1995}, which is based on creation of induced dipoles in the solvent molecules by the oscillating light field during the first few hundred femtoseconds after the pump pulse.

\section{Conclusions}

In summary, we have shown how XCCA can be applied as a model independent approach to study the structural symmetries and their timescale of a disordered sample of solvated photo-excited PtPOP molecules with a preferred photo-induced orientation.
We revealed two time scales that may be attributed to internal structural changes on short time scales and rotational dephasing at longer times.
Analysis of $n=2$ angular Fourier component of CCF enhance structural information, which is otherwise difficult to access in a conventional SAXS approach.
In an ultrafast experiment with a smaller X-ray beam and at high flux facility it might be possible to detect higher order scattering terms necessary for studies of dynamics of internal symmetries of the molecules.
We believe that the technique presented here can be widely used in SAXS/WAXS experiments to enhance structural information from a disordered sample of molecules, proteins or biomolecules and to reveal hidden symmetries and their time evolution in a model independent approach.

\begin{acknowledgments}
We acknowledge E. Weckert for fruitful discussions and support of the project and L. Gelisio for a careful reading of the manuscript.
RPK and IAV acknowledge support from the Helmholtz Association’s Initiative and Networking Fund and the Russian Science Foundation (project No. 18-41-06001). The authors would like to acknowledge Henrik T. Lemke, Silke Nelson, Mike Glownia, T.B. van Driel and Morten Christensen for fruitful scientific discussions. The DTU-affiliated authors would like to gratefully acknowledge DANSCATT for funding the beam time efforts. MMN, KBM, and EB gratefully acknowledge support from the Danish Council For Independent Research under grant no. DFF 4002-00272B. MMN and KBM gratefully acknowledge support from the Independent Research Fund Denmark under grant no. 8021-00347B. Use of the Linac Coherent Light Source (LCLS), SLAC National Accelerator Laboratory, is supported by the U.S. Department of Energy, Office of Science, Office of Basic Energy Sciences under Contract No. DE-AC02-76SF00515.
\end{acknowledgments}

\nocite{*}
\bibliography{PtPOP_references}

\end{document}